\documentclass[a4paper,12pt]{article}
\usepackage{amsmath}
\usepackage{amssymb}
\usepackage{amsthm}
\usepackage{url}
\usepackage[dvips]{graphicx}

\numberwithin{equation}{section}

\DeclareMathOperator{\tah}{th}
\DeclareMathOperator{\Rer}{Re}
\DeclareMathOperator{\Imi}{Im}

\begin{document}

\baselineskip=22pt

\author{V.D. Lakhno}
\title{Pekar Ansatz and the Strong Coupling Problem in Polaron Theory.}
\date{Institute of Mathematical Problems of Biology, RAS}
\maketitle

\baselineskip=22pt

\begin{abstract}
A detailed consideration is given to the translation-invariant
theory of Tulub polaron constructed without the use of Pekar
ansatz. A fundamental result of the theory is that the value of
the polaron energy is lower than that obtained on the basis of
Pekar ansatz which was considered as an asymptotically exact
solution in the strong coupling limit. In the case of bipolarons
the theory yields the best values of the coupling energy and
critical parameters of their stability. Numerous physical
consequences of the existence of translation-invariant polarons
and bipolarons are discussed.
\end{abstract}


Table of contents

1. Pekar ansatz.

2. Coordinate-free Hamiltonian. Weak coupling.

3. Coordinate-free Hamiltonian. General case.

4. Weak coupling limit in Tulub theory.

5. Strong coupling.

6. TI-bipolaron.

7. Functional of the ground state. Tulub ansatz.

8. Discussion of the completeness of Tulub theory.

9. Consequences of the existence of translation-invariant polarons and bipolarons.

10. Conclusive remarks.

Appendix 1.

Appendix 2.

Appendix 3.

References.

\section{Introduction. Pekar ansatz.}

As is known, polaron theory was among the first to describe
the interaction between a particle and a quantum field.
Various aspects of the polaron theory are presented in
numerous reviews and books \cite{1}-\cite{10n-Lak-Kash}.
Being non-relativistic, the theory does not contain any divergencies
and for more than sixty years has been a testing area for
approbation of various methods of the quantum field theory.
Though no exact solution of the polaron problem has been
found up to now, it has been believed that the properties
of the ground state are known in detail. This primarily
refers to the limit cases of weak and strong coupling.
A solution in the weak coupling limit was given by Froehlich
~\cite{9}, and that in the strong coupling one was found by Pekar
 \cite{1},\cite{10}. By now rather an exact solution has been
obtained for the energy of the polaron ground state in the weak
coupling limit \cite{11},\cite{12} :
\begin{equation}\label{1.1}
    E_0=-\left(\alpha+ 0.0159196220\alpha^2 + 0.000806070048\alpha^3+
    \cdot\cdot\cdot\right)\hbar\omega_0\,,
\end{equation}
where  $\hbar\omega_0$  is the energy of an optical phonon, $\alpha$
is the constant of electron-phonon coupling.

A solution of the problem in the opposite strong coupling limit
was given by Pekar on the assumption that the wave function $\Psi$
of the electron + field system has the form:
\begin{equation}\label{1.2}
    \Psi(r,q_1,...,q_i,...)=\psi(r)\Phi(q_1,...,q_i,...)\,,
\end{equation}
where $\psi(r)$   is the electron’s wave function depending only on
the electron coordinates,  $\Phi$ -- is the wave function of the field
depending only on the field coordinates. Pekar himself \cite{1}
considered ansatz  \eqref{1.2} ) to be an approximate solution.
In the pioneer works by Bogolyubov and Tyablikov
\cite{13},\cite{14} it was shown that in a consistent translation-invariant
theory the use of ansatz  \eqref{1.2}
(for decomposed coordinates introduced in  \cite{13},\cite{14} ) gives the
same results for the polaron ground state energy as the semiclassical Pekar
theory does \cite{1},\cite{10}. With the use of \eqref{1.2} the value of the
ground state energy has been found to a high precision.
According to  \cite{15},\cite{16}  it is equal to::
\begin{equation}\label{1.3}
    E=\left(-0.108513\alpha^2-2.836\right)\hbar\omega _0\,.
\end{equation}

The idea that Pekar ansatz \eqref{1.2} is an exact solution of
the strong coupling polaron problem was completely established after
publication of \cite{17} where asymptotics \eqref{1.3} was strictly
proved by path integral method, i.e. without the use of ansatz \eqref{1.2}
(see also review \cite{18}).

Before the publication of paper \cite{17} many attempts were made
to improve the strong coupling theory \cite{19}-\cite{24}.
The reason why Pekar ansatz caused the feeling of disappointment
was translation invariance of the initial polaron Hamiltonian.
When ansatz for the wave function $\psi(r)$ \eqref{1.2}
is used the wave equation
has a localized solution. The electron is localized in a potential
polarization well induced by it. In other words, the solution obtained
does not possess the symmetry of the initial Hamiltonian.
Self-trapping of the electron in the localized potential well leads
to a spontaneous breaking of the system’s symmetry. Attempts to restore
the initial symmetry were based on the use of degeneration of the
system with broken symmetry. Since in a homogeneous and isotropic
medium nothing should depend on the position of the polaron well
center $r_0$, one can "spread"\ the initially localized solution
over all the positions of the polaron potential well by choosing
the wave function in the form of a linear combination in all the
positions of the well.

In the most consistent form this program was carried out in \cite{22}.
With this end in view for the wave function which is an eigen
function of the total momentum, the authors used a superposition
of plain waves corresponding to the total momentum multiplied by
wave functions obtained from \eqref{1.2} to which a translations
operator is applied. In other words, they took an appropriate
superposition with respect to all the positions of the polaron well $r_0$.
The main result of paper \cite{22} is that calculation of the polaron
ground state energy with such a delocalized function yields the
same value as calculations with localized function \eqref{1.2} do.
The authors of \cite{22} also reproduced the value of the polaron
mass which was earlier obtained by Landau and Pekar ~\cite{25} on the
assumption that polaron moves in medium in the localized state  \eqref{1.2}.
The results derived in \cite{22} were an important step in resolving
the contradiction between the requirement that the
translation-invariant wave function be delocalized while the
wave function of the self-trapped state be localized.

Notwithstanding the success achieved with this approach it
cannot be considered fully adequate since it has quite a few
inconsistencies. They follow from the very nature of the
semiclassical description used. Indeed, the superposition
constructed in \cite{22} on the one hand determines the polaron
delocalized state, but on the other hand, without changing
this state, one can measure its position and find out a localized
polaron well with an electron localized in it. The reason of this
paradox is a classical character of the polaron well in the strong
coupling limit and, as a consequence, commutation of the total
momentum operator with the position of the polaron well ~$r_0$.
\footnote{At the rise of quantum mechanics, the founders of the science were fully aware of the difficulties arising here. Thus, for example, in \cite{26n-Bete} Bethe notices that for a proper quantum-mechanical description of an interaction between a field and particles, quantizing of the field is required, i.e. quantum theory of the field: "… The fact is that, when quantizing mechanical parameters (coordinates and momenta) one should also quantize the associated fields. Otherwise, as Bohr and Rosenfeld showed \cite{27n-Bohr}, an imaginary experiment can be suggested which consists in simultaneous measurement of the coordinate and the momentum of a particle from examination of the field induced by it. This contravenes Heisenberg's Uncertainty Principle."}

To remedy this defect some approaches were suggested in which the
quantity $r_0(r,q_1,...,q_i,...)$ which is not actually
an additional degree of freedom
was considered to be that though with some additional constraints.
Discussion of these challenges associated with solution of the
problem of introducing collective coordinates is given in review \cite{26}.
Since the results obtained by introducing collective coordinates
into the polaron theory are polemical it seems appropriate to
describe strict results of the translation-invariant theory
without recourse to the concept of collective coordinates.
The aim of this review is to present an approach used in the
strong coupling limit which does not use Pekar ansatz.

A solution possessing these properties in the case of a strong
coupling polaron was originally found by Tulub \cite{27},\cite{28}.
During nearly half a century the result obtained in \cite{27},\cite{28}
was not  recognized by specialists working in
the field of polaron theory. The reason why the importance of
that result was not  appreciated was an improper
choice of the probe wave function in \cite{28} to estimate the ground state.
As a result the ground state energy was found in \cite{28}
to be $E_0=-0.105\alpha^2\hbar\omega_0$
which is larger than in \eqref{1.3}. An appropriate choice of the wave
function has been made quite recently in paper \cite{29}.
This has yielded a lower than in \eqref{1.3} value of the polaron
ground state energy equal to $E_0=-0.125720\alpha^2\hbar\omega_0$.
Hence, actually we have to do
with inapplicability of adiabatic approximation in the case of
a polaron, though it is fundamental for solid state physics.

In this review we present the main points of the translation-invariant
polaron (TI-polaron) theory. In \S2 Heisenberg canonical transformation
is used to introduce a coordinate-free Pekar – Froehlich Hamiltonian
which forms the basis for translation-invariant description.
With the use of Lee-Low-Pines wave function the weak coupling
limit is reproduced.

In \S3 we present the general translation-invariant Tulub theory
valid for any values of the electron-phonon coupling constant.
In view of primary importance of Tulub approach the material
is presented in greater detail than in the original paper so
that a reader may reproduce and check the results obtained
if desired. A general expression is obtained for the TI-polaron
energy for an arbitrary coupling strength.

In \S4 a limit case of weak coupling is considered. It is shown
that the general expression for the polaron energy derived in
the previous section reproduces the weak coupling limit to a
high precision.

In \S5  a limit case of strong coupling is dealt with. It is shown
that in this limit case the polaron ground state energy has a
lower value as compared to that obtained on the basis
of Pekar ansatz.

In \S6 the translation-invariant polaron theory is generalized
to the case of a bipolaron. The bipolaron ground state energy
obtained here is much lower than that yielded by the best
variational calculations with the use of Pekar ansatz.
The calculation results are used to explain the
high-temperature superconductivity by using TI-bipolarons.

In \S7 the energy of TI-polaron and bipolaron is derived in
an alternative way. The approach presented there enables
one to get an explicit form of wave functions for a polaron
and bipolaron. The results obtained suggest that for all
the values of the electron-phonon coupling constant the
wave functions derived describe delocalized states.
Hence it is shown that in the strong coupling limit
Pekar ansatz is not fulfilled and no transition
to the self-trapped (i.e. localized) state with broken
symmetry takes place.

The results obtained radically change the visions of polarons
and bipolarons and in the general case cast some doubt on the
concept of self-localized states in condensed systems.
Recently an open discussion on the completeness of Tulub
theory has taken place. The main points and results of the
discussion are presented in \S8.

In \S9 we discuss some problems and consequences emerging from
the existence of TI-polarons and bipolarons.
Of special practical interest are results concerned
with their superconducting properties.

\S10  is devoted to discussion of some fundamental problems
of the strong coupling theory which are still to be solved.

In Appendices 1-3 we give proofs of some important statements
basic to the approach under consideration.

\section{Coordinate-free Hamiltonian. Weak coupling.}

Let us proceed from Pekar-Froehlich Hamiltonian:
\begin{equation}\label{2.1}
    H=-\frac{\hbar^2}{2m}\Delta_r+\sum_{k}V_k\left(a_ke^{ikr}+a_k^{+}e^{-ikr}\right)+\sum_{k}\hbar\omega_k^{0}a_k^{+}a_k\,,
\end{equation}
where $a_k^{+}$, $a_k$ are operators of the birth and annihilation
of the field quanta with energy $\hbar\omega_k^{0}=\hbar\omega_0$, $m$
is the electron effective mass,
$V_k$ is a function of the wave vector $k$.

Interest in the study of this Hamiltonian is also provoked by the fact that, as distinct from many model Hamiltonians considered in the condensed matter theory, Pekar-Froehlich Hamiltonian \eqref{2.1} in the limit of long waves asymptotically exactly describes the behavior of a non-relativistic electron in a continuous polar medium.

Electron coordinates can be excluded from \eqref{2.1}
via Heisenberg transformation \cite{30}:
\begin{equation}\label{2.2}
    S_1=\exp\left\{\frac{i}{\hbar}\left(\vec{P}-\sum_{k}\hbar\vec{k}a_k^{+}a_k\right)\vec{r}\right\}\,,
\end{equation}
where $\vec{P}$ is the total momentum of the system.
Application of $S_1$ to the field operators yields:
\begin{equation*}
    S_1^{-1}a_kS_1=a_ke^{-ikr}\,,\ \ \
    S_1^{-1}a_k^{+}S_1=a_k^{+}e^{ikr}\,.
\end{equation*}
Accordingly the transformed Hamiltonian $\tilde{H}=S_1^{-1}HS_1$ takes on the form:
\begin{equation}\label{2.3}
    \tilde{H}=\frac{1}{2m}\left(\vec{P}-\sum_{k}\hbar
    \vec{k}a_k^{+}a_k\right)^2+\sum_{k}V_k(a_k+a_k^{+})+\sum_{k}\hbar\omega_k^{0}a_k^{+}a_k\,.
\end{equation}
Since Hamiltonian \eqref{2.3} does not contain electron coordinates,
it is obvious that solution of the polaron problem obtained on the
basis of \eqref{2.3} is translation-invariant. Lee, Low and Pines \cite{31}
studied the ground state \eqref{2.3} with the probe wave function $|\Psi\rangle_{LLP}$ :
\begin{equation}\label{2.4}
    |\Psi\rangle_{LLP}=S_2|0\rangle\,,
\end{equation}
where:
\begin{equation}\label{2.5}
    S_2=\exp\left\{\sum_{k}f_k(a_k^{+}-a_k)\right\}\,,
\end{equation}
$f_k$  are variational parameters having the meaning of the value
of displacement of the field oscillators from their equilibrium positions,
$|0\rangle$  is the vacuum wave function. The quantity $f_k$ in $S_2$ \eqref{2.5}
is determined by minimization of energy $E=\langle0|S_2^{-1}\tilde{H}S_2|0\rangle$,
which for $P=0$  yields:
\begin{equation}\label{2.6}
    E=2\sum_{k}f_kV_k+\frac{\hbar^2}{2m}\left[\sum_{k}\vec{k}f_k^2\right]^2+
    \sum_{k}\frac{\hbar^2k^2}{2m}f_k^2+\sum_{k}\hbar\omega_k^{0}f_k^2\,,
\end{equation}
\begin{equation}\label{2.7}
    f_k=-\frac{V_k}{\hbar\omega_k^{0}+\hbar^2k^2/2m}\,.
\end{equation}

In the case of an ionic crystal:
\begin{equation}\label{2.8}
    V_k=\frac{e}{k}\sqrt{\frac{2\pi\hbar\omega_0}{\tilde{\varepsilon}V}}=\frac{\hbar\omega_0}{ku^{1/2}}\left(\frac{4\pi\alpha}{V}\right)^{1/2}\,,\
    u=\left(\frac{2m\omega_0}{\hbar}\right)^{1/2}\,,\
    \alpha=\frac{1}{2}\frac{e^2u}{\hbar\omega_0\tilde{\varepsilon}}\,,\
    \tilde{\varepsilon}^{-1}=\varepsilon_{\infty}^{-1}-\varepsilon_0^{-1}\,,
\end{equation}
where $e$ is an electron charge, $\varepsilon_{\infty}$ and $\varepsilon_0$
are high-frequency and static dielectric permittivities,
$\alpha$ is a constant of electron-phonon coupling.
With substitution of \eqref{2.8} into \eqref{2.6}, \eqref{2.7} the ground state energy
becomes $E=-\alpha\hbar\omega_0$, which is the energy of a weak coupling polaron in the first order
with respect to $\alpha$.

A solution of the problem of transition to the strong
coupling case in coordinate-free Hamiltonian \eqref{2.3}  was found on
the basis of the general translation-invariant theory constructed
in Tulub’s work \cite{28}. The main points of this theory are given in
the next section.

\section{Coordinate-free Hamiltonian. General case.}

To construct the general translation-invariant theory in works
of \cite{27}, \cite{28} was used a canonical transformation of Hamiltonian \eqref{2.3}
with the use of operator $S_2$ \eqref{2.5} which leads to a shift of the field operators:
\begin{equation}\label{3.1}
    S_2^{-1}a_kS_2=a_k+f_k\,,\ \ S_2^{-1}a_k^{+}S_2=a_k^{+}+f_k\,.
\end{equation}
The resultant Hamiltonian $\tilde{\tilde{H}}=S_2^{-1}\tilde{H}S_2$
has the form:
\begin{equation}\label{3.2}
    \tilde{\tilde{H}}=H_0+H_1\,,
\end{equation}
where:
\begin{equation}\label{3.3}
    H_0=\frac{\vec{P}^2}{2m}+2\sum_{k}V_kf_k +
    \sum_{k}\left(\hbar\omega_k^{0}-\frac{\hbar\vec{k}\vec{P}}{m}\right)f_k^2
    + \frac{1}{2m}\left(\sum_{k}\vec{k}f_k^2\right)^2+
    \mathcal{H}_0\,,
\end{equation}
\begin{equation}\label{3.4}
    \mathcal{H}_0=\sum_{k}\hbar\omega_ka_k^{+}a_k +
    \frac{1}{2m}\sum_{k,k'}\vec{k}\vec{k}'f_kf_{k'}\left(a_ka_{k'}+a_k^{+}a_{k'}^{+}+a_k^{+}a_{k'}+a_{k'}^{+}a_k\right)\,,
\end{equation}
\begin{equation}\label{3.5}
    \hbar\omega_k=\hbar\omega_k^{0}-\frac{\hbar\vec{k}\vec{P}}{m}+\frac{\hbar^2k^2}{2m}+\frac{\hbar\vec{k}}{m}\sum_{k'}\hbar\vec{k}'f_{k'}^2\,.
\end{equation}
Hamiltonian $H_1$ contains terms linear, triple and quadruple
in the birth and annihilation operators. With an appropriate
choice of the wave function diagonalizing quadratic form \eqref{3.4}
mathematical expectation $H_1$ becomes zero (Appendix 1).
In what follows we believe that $\hbar=1$,
$\omega_0=1$, $m=1$. To transform
$\mathcal{H}_0$ to a diagonal form we put:
\begin{equation}\label{3.6}
    q_k=\frac{1}{\sqrt{2\omega_k}}(a_k+a_k^{+})\,,\ \
    p_k=-i\sqrt{\frac{\omega_k}{2}}(a_k-a_k^{+})\,,\ \
    \vec{z}_k=\vec{k}f_k\sqrt{2\omega_k}\,.
\end{equation}
With the use of \eqref{3.6} expression \eqref{3.4} is written as:
\begin{equation}\label{3.7}
    \mathcal{H}_0=\frac{1}{2}\sum_{k}(p_k^{+}p_k+\omega_k^2q_k^{+}q_k)
    +
    \frac{1}{2}\left(\sum_{k}\vec{z}_kq_k\right)^2-\frac{1}{2}\sum_{k}\omega_k\,.
\end{equation}
This yields \eqref{3.7} the following motion equation for operator $q_k$ :
\begin{equation}\label{3.8}
    \ddot{q}_k+\omega_k^2q_k=-\vec{z}_k\sum_{k'}\vec{z}_{k'}q_{k'}\,.
\end{equation}
Let us search for a solution of system \eqref{3.8} in the form:
\begin{equation}\label{3.9}
    q_k(t)=\sum_{k'}\Omega_{kk'}\xi_{k'}(t)\,,\ \
    \xi_{k}(t)=\xi_{k}^{0}e^{i\nu_{k}t}\,.
\end{equation}
As a result we express matrix $\Omega_{kk'}$ as:
\begin{equation}\label{3.10}
    (\nu_{k'}^2-\omega_k^2)\Omega_{kk'}=\vec{z}_k\sum_{k''}\vec{z}_{k''}\Omega_{k''k'}\,.
\end{equation}
Let us consider determinant of this system which is derived
by replacing the eigenvalues
$\nu_k^2$ in \eqref{3.10} with the quantity $s$ which can differ from $\nu_k^2$.
The determinant of this system will be
\begin{equation}\label{3.11}
    \det\left|(s-\omega_k^2)\delta_{kk'}-\vec{z}_k\vec{z}_{k'}\right|=\prod_{k}(s-\nu_k^2)\,.
\end{equation}
On the other hand, according to \cite{32}\footnote{In Wentzel's work $z_k$ is not a vector function, but a scalar one, therefore a "cube"\,in \eqref{3.12} is lacking. Generalization to the vector case is given in \cite{27}.}:
\begin{equation}\label{3.12}
    \det\left|(s-\omega_k^2)\delta_{kk'}-\vec{z}_k\vec{z}_{k'}\right|=\prod_{k}(s-\omega_k^2)
    \left(1-\frac{1}{3}\sum_{k'}\frac{\vec{z}_{k'}^2}{s-\omega_{k'}^2}\right)^3\,.
\end{equation}
It is convenient to introduce the quantity $\Delta(s)$ :
\begin{equation}\label{3.13}
    \Delta(s)=\prod_{k}(s-\nu_k^2)/\prod_{k}(s-\omega_k^2)\,.
\end{equation}
With the use of \eqref{3.11} and \eqref{3.12}
$\Delta(s)$ is expressed as:
\begin{equation}\label{3.14}
    \Delta(s)=\left(1-\frac{1}{3}\sum_{k}\frac{\vec{z}_{k}^2}{s-\omega_k^2}\right)^3\,.
\end{equation}
From \eqref{3.11}, \eqref{3.12} follows that
the frequencies $\nu_k$ renormalized by interaction  are determined
by a solution to the equation:
\begin{equation}\label{3.15}
    \Delta(\nu_k^2)=0\,.
\end{equation}
The change in the system’s energy $\Delta E$ caused by the
electron-field interaction is equal to:
\begin{equation}\label{3.16}
    \Delta E=\frac{1}{2}\sum_{k}(\nu_k-\omega_k)\,.
\end{equation}
To express the quantity $\Delta E$ via $\Delta(s)$ we use Wentzel
approach \cite{32}. Following \cite{32} we write down the identity equation:
\begin{multline}\label{3.17}
    \sum_{k}\left\{f(\nu_k^2)-f(\omega_k^2)\right\}=\frac{1}{2{\pi}i}\oint\limits_{C}dsf(s)\sum_{k}\left(\frac{1}{s-\nu_k^2}-\frac{1}{s-\omega_k^2}\right)=\\
    =\frac{1}{2\pi i}\oint\limits_{C}dsf(s)\frac{d}{ds}\ln\Delta(s)=-\frac{1}{2\pi
    i}\oint\limits_{C}dsf'(s)\ln\Delta(s)\,,
\end{multline}
where integration is carried out over the contour presented in Fig.1
\begin{center}

\includegraphics{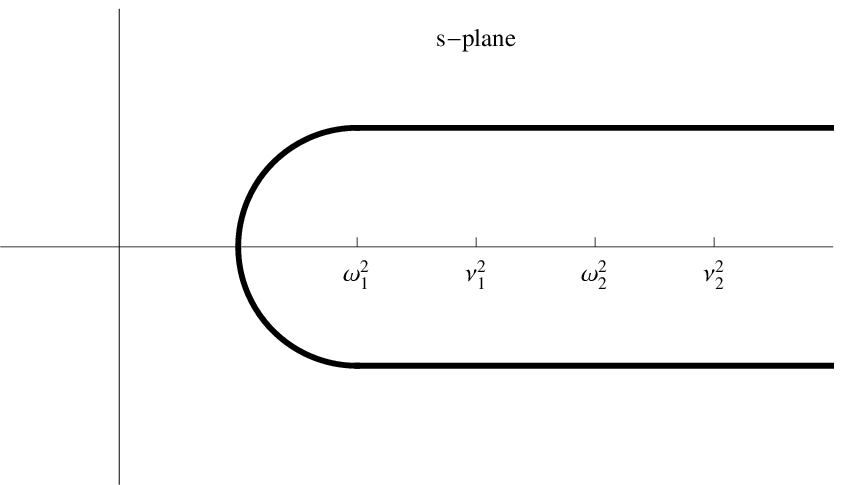}

Fig. 1. Contour $C$.
\end{center}

Taking $f(s)=\sqrt{s}$ we get:
\begin{equation}\label{3.18}
    \Delta E=\frac{1}{2}\sum_{k}(\nu_k-\omega_k)=-\frac{1}{8\pi
    i}\oint\limits_{C}\frac{ds}{\sqrt{s}}\ln\Delta(s)\,.
\end{equation}
Turning in \eqref{3.14} from summing up to integration with the use of the relation:
\begin{equation*}
    \sum_{k}=\frac{1}{(2\pi)^3}\int d^3k
\end{equation*}
in a continuous case, using for $\vec{z}_k$ expression
\eqref{3.14} for  $\Delta(s)$ we obtain:
\begin{equation}\label{3.19}
    \Delta(s)=D^3(s)\,,\ \
    D(s)=1-\frac{2}{3(2\pi)^3}\int\frac{k^2f_k^2\omega_k^2}{s-\omega_k^2}d^3k\,.
\end{equation}
As a result the total energy of the electron is:
\begin{equation}\label{3.20}
    E=\Delta E + 2\sum_{k}V_kf_k + \sum_{k}f_k^2\omega_k^0\,.
\end{equation}
The results obtained here are general and valid for various
polaron models (i.e. any functions $V_k$ and $\omega_k^0$). Below we
consider limit cases of weak and strong coupling which
follow from general expression \eqref{3.20} on the assumption
that $\vec{P}=0$.

Notice that for $\vec{P}\neq0$, according to \cite{27}, expression \eqref{3.20} takes the form:

\begin{equation*}
    E=\frac{P^2}{2m}+\Delta E(\vec{P}) + 2\sum_{k}V_kf_k + \sum_{k}\left(\hbar\omega_k^{0}-\frac{\hbar\vec{k}\vec{P}}{m}\right)f_k^2
    + \frac{1}{2m}\left(\sum_{k}\vec{k}f_k^2\right)^2\,,
\end{equation*}

\begin{equation*}
    \Delta E(\vec{P})=-\frac{1}{8\pi i}\oint\limits_{C}\frac{ds}{\sqrt{s}}\ln\prod_{i=1}^{3}D^i(s)\,,
\end{equation*}

\begin{equation*}
D^i(s)=1-\sum_{k}\frac{\left(z_k^i\right)^2}{s-\omega_k^2}\,,
\end{equation*}
where $z_k^i$ -- $i$-th component of the vector $\vec{z}_k$. Functions $f_k$, $\omega_k$ and $z_k$ are depending on $|\vec{k}|$ and on $(\vec{k}\vec{P})$.

\section{Weak coupling limit in Tulub theory.}

Quantities $f_k$ in the expression for the total energy
$E$ should be found from the minimum condition:
$\delta E/\delta f_k=0$ which yields the following
integral equation for $f_k$:
\begin{equation}\label{4.1}
    f_k=-V_k/(1+k^2/2\mu_k)\,,\ \ \mu_k^{-1}=\frac{\omega_k}{2\pi
    i}\oint\limits_{C}\frac{ds}{\sqrt{s}}\frac{1}{(s-\omega_k^2)D(s)}\,.
\end{equation}
In the case of weak coupling $\alpha\to0$ and equations \eqref{4.1}
can be solved with the use of perturbation theory.
In a first approximation as $\alpha\to0$, $D(s)=1$ and $\mu_k^{-1}$ is equal to:
\begin{equation}\label{4.2}
    \mu_k^{-1}=\frac{\omega_k}{2\pi
    i}\oint\limits_{C}\frac{ds}{\sqrt{s}}\frac{1}{(s-\omega_k^2)}=1\,.
\end{equation}
Accordingly, $f_k$ from \eqref{4.1} is written as:
\begin{equation}\label{4.3}
    f_k=-V_k/(1+k^2/2)\,.
\end{equation}
The quantity $\Delta E$ involved in the total energy takes on the form:
\begin{equation}\label{4.4}
    \Delta E=-\frac{3}{8\pi
    i}\oint\limits_{C}\frac{ds}{\sqrt{s}}\ln D(s)\,,\ \ \ln
    D(s)=-\frac{2}{3(2\pi)^3}\int\frac{k^2f_k^2\omega_k}{s-\omega_k^2}d^3k\,.
\end{equation}
With the use of \eqref{4.3} integrals involved in \eqref{4.4}
are found to be: $\Delta E=(\alpha/2)\hbar\omega _0$.
Having calculated the rest of the terms involved in expression
\eqref{3.20} we get the first term of the expansion of polaron
total energy in the coupling constant  $\alpha$ : $E=-\alpha\hbar\omega _0$.

In papers \cite{27}, \cite{33}, \cite{34} a general scheme of calculating
the higher terms of expansion in $\alpha$ was
developed. In particular, the eigen energy and effective
mass were found to be \cite{34}:
\begin{equation}\label{4.5}
\begin{split}
    E&=-(\alpha+0.01592\alpha^2)\hbar\omega _0\,, \\
    m^*&=\left(1+\frac{\alpha}{6}+0.02362\alpha^2\right)m\,.
\end{split}
\end{equation}
Hence within the accuracy of the terms $O(\alpha^3)$ the polaron energy
expression calculated within Tulub approach with the use of
perturbation theory coincides with exact result \eqref{1.1} (see \S7).

\section{Strong coupling.}

The case of strong coupling is much more complicated. To reveal the
character of the solution in the strong coupling region let us start
with considering the analytical properties of the function $D(s)$ in the form:
\begin{equation}\label{5.1}
    D(s)=D(1)+\frac{s-1}{3\pi^2}\int\limits_{0}^{\infty}\frac{k^4f_k^2\omega_k
    dk}{(\omega_k^2-1)(\omega_k^2-s)}\,,
\end{equation}
where $D(1)$ is the value of $D(s)$ for $s=1$:
\begin{equation}\label{5.2}
    D(1)=1+Q\equiv1+\frac{1}{3\pi^2}\int\limits_{0}^{\infty}\frac{k^4f_k^2\omega_k}{\omega_k^2-1}dk\,.
\end{equation}
From \eqref{3.19} also follows that:
\begin{equation}\label{5.3}
    D(s)=1-\frac{1}{3\pi^2}\int\limits_{0}^{\infty}\frac{\omega_kk^4f_k^2}{s-\omega_k^2}dk\,.
\end{equation}
Function $D(s)$, being a function of a complex variable $s$,
has the following properties:
1) $D(s)$ has a crosscut along the real axis from $s=1$ to $\infty$
and has no other peculiarities;
2) $D^*(s)=D(s^*)$;
3) as $s\to\infty$ $sD(s)$ increases not slower than $s$.
These properties enable us to present the function $[(s-1)D(s)]^{-1}$
in the form (Appendix 2):
\begin{equation}\label{5.4}
    \frac{1}{(s-1)D(s)}=\frac{1}{2\pi
    i}\oint\limits_{C+\rho}\frac{ds'}{(s'-s)(s'-1)D(s')}\,,
\end{equation}
where contour $C+\rho$ is shown in Fig. 2:
\begin{center}

\includegraphics{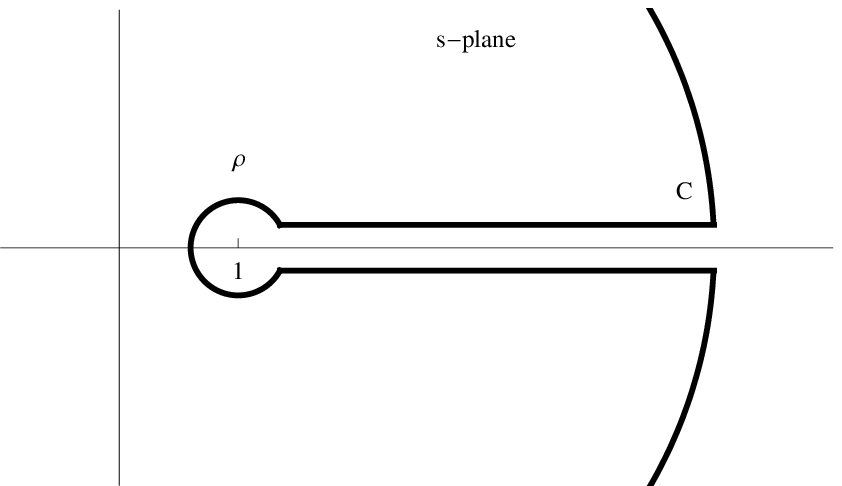}

Fig. 2. Contour $C+\rho$.
\end{center}

 The integrand function in \eqref{5.4} has a pole at $s'=1$  and a section
from $s'=1$ to ${s'=\infty}$. Having performed integration in \eqref{5.4} along
the upper and bottom sides of the crosscut we get the
following integral equation for $D^{-1}(s)$ :
\begin{equation}\label{5.5}
    \frac{1}{D(s)}=\frac{1}{1+Q}+\frac{s-1}{3\pi^2}\int\limits_{0}^{\infty}\frac{k^4f_k^2\omega_k
    dk}{(s-\omega_k^2)(\omega_k^2-1)|D(\omega_k^2)|^2}\,.
\end{equation}
With the use of integration by parts expression \eqref{3.18} for $\Delta E$ can be written as:
\begin{equation}\label{5.6}
    \Delta E=\frac{1}{2\pi^2}\int\limits_{0}^{\infty}dk
    k^4f_k^2\omega_k\frac{1}{2\pi
    i}\oint\limits_{C}\frac{\sqrt{s}}{(s-\omega_k^2)^2}\frac{1}{D(s)}ds\,.
\end{equation}
From \eqref{5.5}, \eqref{5.6} we have:
\begin{equation}\label{5.7}
     \Delta
     E=\frac{1}{2\pi^2}\int\limits_{0}^{\infty}\frac{k^4f_k^2dk}{2(1+Q)}+\frac{1}{12\pi^4}\int\limits_{0}^{\infty}\int\limits_{0}^{\infty}
     \frac{k^4f_k^2p^4f_p^2\omega_p(\omega_k\omega_p+\omega_k(\omega_k+\omega_p)+1)}{(\omega_k+\omega_p)^2(\omega_p^2-1)|D(\omega_p^2)|^2}dpdk\,.
\end{equation}
Equation $\mu_k^{-1}$ \eqref{4.1}, according to \eqref{5.5}, can be presented in the form:
\begin{equation}\label{5.8}
    \mu_k^{-1}=\frac{1}{1+Q}+
    \frac{1}{3\pi^2}\int\limits_{0}^{\infty}\frac{p^4f_p^2(\omega_k\omega_p+1)dp}{(\omega_p^2-1)(\omega_k+\omega_p)|D(\omega_p^2)|^2}\,.
\end{equation}
Equations \eqref{4.1}, \eqref{5.8} for finding $f_k$ as well as expressions
\eqref{3.20}, \eqref{5.7} for calculating polaron energy are very complicated
and their exact solution can hardly be obtained. To calculate approximately
the energy $E$  given by \eqref{3.20}, \eqref{5.7} in \cite{28} a direct variational principle
was used. For the probe function, the author used Gaussian function of the form:
\begin{equation}\label{5.9}
    f_k=-V_k\exp(-k^2/2a^2)\,,
\end{equation}
where $a$ is a variable parameter, besides, as can be seen in the case
of strong coupling, $a\gg1$. Substitution of \eqref{5.9} into \eqref{3.19} yields for
real and imaginary parts of $D(s)$ (see Appendix 3):
\begin{equation}\label{5.10}
    \begin{split}
    &\Rer D(\omega_k^2)=1+\lambda v(y)\,, \ \Imi D(\omega_k^2)=k^3f_k^2/6\pi\,,\\
    &v(y)=1-ye^{-y^2}\int\limits_{0}^{y}e^{t^2}dt -
    ye^{y^2}\int\limits_{y}^{\infty}e^{-t^2}dt\,,\\
    &\lambda=4\alpha a/3\sqrt{2\pi}\,,\ \ y=k/a\,.
\end{split}
\end{equation}
In the limit of strong coupling ($\alpha\gg1$) the expression for energy $E$
given by \eqref{3.20} with the use of \eqref{5.7} takes on the form:
\begin{equation}\label{5.11}
    E=\frac{3}{16}a^2\left[1+q\left(\frac{1}{\lambda}\right)\right]
    - \frac{\alpha
    a}{\sqrt{\pi}}\left(2-\frac{1}{\sqrt{2}}\right)\,,
\end{equation}
\begin{equation}\label{5.12}
    q\left(\frac{1}{\lambda}\right)=\frac{2}{\sqrt{\pi}}\int\limits_{0}^{\infty}\frac{e^{-y^2}(1-\Omega(y))dy}{(1/\lambda+v(y))^2+\pi y^2e^{-2y^2}/4}\,,
\end{equation}
\begin{equation*}
    \Omega(y)=2y^2\left\{(1+2y^2)ye^{y^2}\int\limits_{y}^{\infty}e^{-t^2}dt
    -y^2\right\}\,.
\end{equation*}
 As $\lambda\rightarrow\infty$, integral \eqref{5.12} has maximum for $y^4=3\lambda/4$, if the function $f_k$ is chosen in the form \eqref{5.9}, however if the actual boundedness of the region of integration with respect to $y$ is taken into account, this peculiarity does not take place (see \S8).

When calculating \eqref{5.12} in paper \cite{28} Tulub assumed that in the strong
coupling limit $1/\lambda=0$. As a result of numerical integration $q(0)$ was found
to be $q(0)=5.75$ whence, varying energy $E$ \eqref{5.11} with respect to $a$ we get:
\begin{equation}\label{5.13}
    E=-0.105\alpha^2\hbar\omega _0\,.
\end{equation}
Comparison of \eqref{5.13} with \eqref{1.3} shows that the value of $E$
obtained for $\alpha\to\infty$  lies higher than the exact value in Pekar
theory \eqref{1.3}. For this reason, until quite recently it was
believed that Tulub theory as applied to a polaron does not
give any new results.

The situation changed radically after publication of \cite{29}.
There it was shown that the choice of the wave function for
minimizing energy \eqref{3.20} in the form \eqref{5.9} is not optimal
since it does not satisfy virial relations. As is shown in
\cite{29}, an appropriate function $f_k$ should contain the
multiplier $\sqrt{2}$  outside the exponent in expression \eqref{5.9}.
Minimization of energy \eqref{3.20} with the optimal probe function yields:
\begin{equation}\label{5.14}
    E=-0.125720\alpha^2 \hbar\omega _0\,.
\end{equation}
Result \eqref{5.14} is fundamental. Above all it means that Pekar ansatz
does not give an exact solution. Though result \eqref{5.14} refers to a
particular case of Pekar–Froehlich Hamiltonian with  $V_k$ given by
\eqref{2.8}, the conceptual conclusion should be valid for all types of
self-localized states. Of special interest is to consider the case
of bipolarons since they can play an important role in superconductivity.

\section{TI-bipolarons.}

Great attention given to polaron problem in recent times is
associated with attempts to explain the superconductivity
phenomenon relying on the mechanism of Bose condensation of
bipolaron gas. In this connection the study of conditions under
which the bipolaron states are stable is of paramount importance.
The theory of large-radius bipolarons which are now considered
to be the best candidates for the role of charged bosons forming
Bose-Einstein condensate with pairing in real space is
considered in detail in reviews \cite{7}, \cite{8}, \cite{35}.

The study of the process of formation of a stable two-electron
state in a crystal, or a bipolaron, generally implies finding
a pairwise interaction between two polarons as a function of
a distance between them \cite{8}. For a large-radius bipolaron,
the region of its existence is bordered on the part of the
coupling constant $\alpha$ by rather a large value of $\alpha_c$ below which
the polaron bound state does not exist. In view of the
requirement that $\alpha_c$ be large, which may not be met in
high-temperature superconductors, some researchers investigated
the contribution of other types of interactions and coupling
symmetries \cite{36}, \cite{37}.

In what follows we will deal with only electron-phonon Pekar–Froehlich
interaction since the approach under consideration can be generalized
to other types of interactions as well. It seems all the more actual
in view of the fact that in recent times some reasoned arguments
have been obtained testifying that electron-phonon interaction
in high-temperature superconductors is strong \cite{38}-\cite{40}.
There are also arguments in favor of the fact that owing to weak
screening of high-frequency optical phonons, the electron-phonon
interaction in high-temperature superconductors is more adequately
described not in the framework of a contact interaction of
Holstein polaron model \cite{41}, but in terms of a long-range interaction
of Froehlich type \cite{42}.

Before the publication of \cite{43}-\cite{45} the lowest values of the energy
of bipolaron states determined by electron-phonon interaction were
obtained in \cite{46}-\cite{48} for $\alpha<8$ and in \cite{48}-\cite{51} for $\alpha>8$.
Attempts to find a translation-invariant solution of the bipolaron
problem by variational methods using direct variation of the wave
function of the two-electron system \cite{35}, \cite{52}, \cite{53} yielded higher
values of the bipolaron ground state energy as compared to those
obtained with the use of a wave function lacking translation
invariance \cite{47}, \cite{48},
\cite{51}, \cite{54}. In this section we present
the results obtained in \cite{43}-\cite{45} for a bipolaron within the
translation-invariant approach.

Let us proceed from Pekar-Froehlich Hamiltonian for a bipolaron
\cite{8} :
\begin{equation}\label{6.1}
    H=-\frac{\hbar^2}{2m}\Delta_{r_1}-\frac{\hbar^2}{2m}\Delta_{r_2}
    + \sum_{k}\hbar\omega_k^{0}a_k^{+}a_k +
    U\left(|\vec{r}_1-\vec{r}_2|\right) + \sum_{k}\left(V_ke^{ikr_1}a_k+V_ke^{ikr_2}a_k +
    H.C.\right)\,,
\end{equation}
\begin{equation*}
    U\left(|\vec{r}_1-\vec{r}_2|\right)=\frac{e^2}{\varepsilon_{\infty}|\vec{r}_1-\vec{r}_2|}\,,
\end{equation*}
where $\vec{r}_1$ and $\vec{r}_2$ are coordinates of the first and
second electrons, respectively, quantity $U$ describes Coulomb
repulsion between the electrons.

In the mass center system Hamiltonian \eqref{6.1} takes on the form:
\begin{equation}\label{6.2}
    H=-\frac{\hbar^2}{2M_e}\Delta_{R}-\frac{\hbar^2}{2\mu_e}\Delta_{r}
    +
    U(|r|)+\sum_{k}\hbar\omega_k^{0}a_k^{+}a_k+\sum_{k}2V_k\cos\frac{\vec{k}\vec{r}}{2}\left(a_ke^{i\vec{k}\vec{R}}+H.C.\right)\,,
\end{equation}
\begin{equation*}
    \vec{R}=(\vec{r}_1+\vec{r}_2)/2\,,\ \
    \vec{r}=\vec{r}_1-\vec{r}_2\,,\ \ M_e=2m\,,\ \ \mu_e=m/2\,.
\end{equation*}
In what follows we will believe that $\hbar=1$, $\omega_k^{0}=1$, $M_e=1$ (accordingly $\mu_e$=1/4).

The coordinates of the mass center $\vec{R}$ can be excluded from
Hamiltonian \eqref{6.2} via Heisenberg canonical transformation:
\begin{equation*}
    S_1=\exp\left\{-i\sum_{k}\vec{k}a_k^{+}a_k\right\}\vec{R}\,,
\end{equation*}
\begin{equation}\label{6.3}
    \tilde{H}=S_1^{-1}HS_1=-2\Delta_{r}+U(|r|)+\sum_{k}a_k^{+}a_k + \sum_{k}2V_k\cos\frac{\vec{k}\vec{r}}{2}(a_k+a_k^{+})
    +\frac{1}{2}\left(\sum_{k}\vec{k}a_k^{+}a_k\right)^2\,.
\end{equation}
From formula \eqref{6.3} follows that the exact solution of the bipolaron
problem is determined by the wave function $\psi(r)$ containing only
relative coordinates $r$ and, therefore, possessing translation invariance.

Averaging of $\tilde{H}$  over $\psi(r)$ leads to the Hamiltonian $\bar{H}$:
\begin{equation}\label{6.4}
    \bar{H}=\frac{1}{2}\left(\sum_{k}\vec{k}a_k^{+}a_k\right)^2+\sum_{k}a_k^{+}a_k+\sum_{k}\bar{V}_k(a_k+a_k^{+})+\bar{T}+\bar{U}\,,
\end{equation}
\begin{equation*}
    \bar{V}_k=2V_k\langle\psi|\cos\frac{\vec{k}\vec{r}}{2}|\psi\rangle\,,\
    \ \bar{U}=\langle\psi|U(r)|\psi\rangle\,,\ \
    \bar{T}=-2\langle\psi|\Delta_r|\psi\rangle\,.
\end{equation*}
Hamiltonian \eqref{6.4} differs from Hamiltonian \eqref{2.3} in that the quantity
$V_k$  in \eqref{2.3} is replaced by $\bar{V}_k$ and constants $\bar{T}$ and $\bar{U}$ are added.
Therefore, repeating the derivation performed in \S3 we express
the bipolaron energy $E_{bp}$ as:
\begin{equation}\label{6.5}
    E_{bp}=\Delta E
    +2\sum_{k}\bar{V}_kf_k+\sum_{k}f_k^2+\bar{T}+\bar{U}\,,
\end{equation}
where $\Delta E$ is given by expression \eqref{5.7}. From \eqref{6.5} we can get expressions
for the bipolaron energy varying $E_{bp}$  with respect to $f_k$ and $\psi$.
Since the equations obtained in this way are difficult to solve,
for actual determining of the bipolaron energy we use
a direct variational approach, assuming \cite{45}:
\begin{equation}\label{6.6}
    \begin{split}
    f_k&=-N\bar{V}_k\exp(-k^2/2\mu)\,,\\
    \psi(r)&=(2/\pi \ell^2)^{3/4}\exp(-r^2/\ell^2)\,,
\end{split}
\end{equation}
where $N$, $\mu$, $\ell$ are variational parameters. For $N=1$, expression \eqref{6.6}
reproduces the results of work \cite{43}, and for
$N=1$ and $\mu\to\infty$ , those of work \cite{44}.

Having substituted \eqref{6.6} into the expression for the total energy
and then minimized the expression obtained with respect to
parameter $N$ we write $E$ as:
\begin{equation}\label{6.7}
    E(x,y;\eta)=\Phi(x,y;\eta)\alpha^2\,,
\end{equation}
\begin{equation*}
    \Phi(x,y;\eta)=\frac{6}{x^2}+\frac{20.25}{x^2+16y}-\frac{16\sqrt{x^2+16y}}{\sqrt{\pi}(x^2+8y)}+\frac{4\sqrt{2/\pi}}{x(1-\eta)}\,.
\end{equation*}
Here $x$, $y$ are variational parameters: $x=\ell\alpha$,
$y=\alpha^2/\mu$, $\eta=\varepsilon_{\infty}/\varepsilon_0$.Let us write $\Phi_{min}$
for the minimum of function $\Phi$ of parameters $x$ and $y$.
Fig.3 shows the dependence of $\Phi_{min}$ on the parameter $\eta$.
Fig.4 demonstrates the dependence of $x_{min}$,
$y_{min}$ on the parameter $\eta$.
\begin{center}

\includegraphics{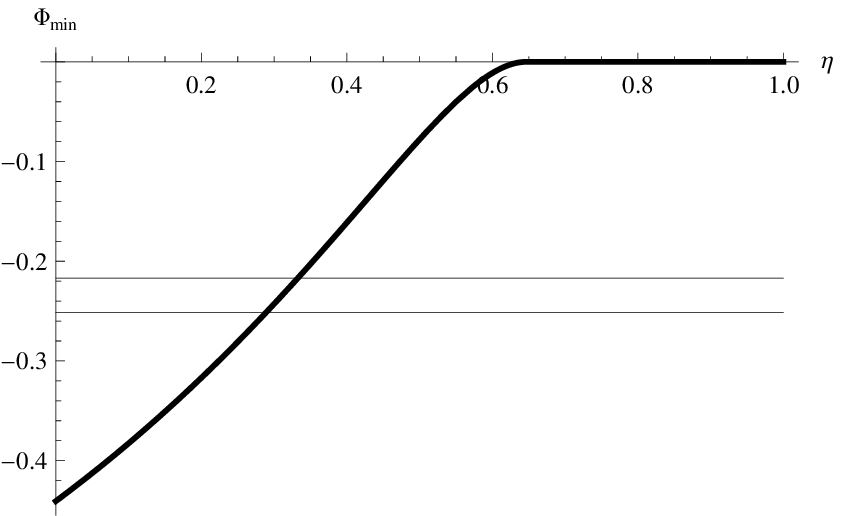}

Fig. 3. Graph $\Phi_{min}(\eta)$.
\end{center}

\begin{center}

\includegraphics{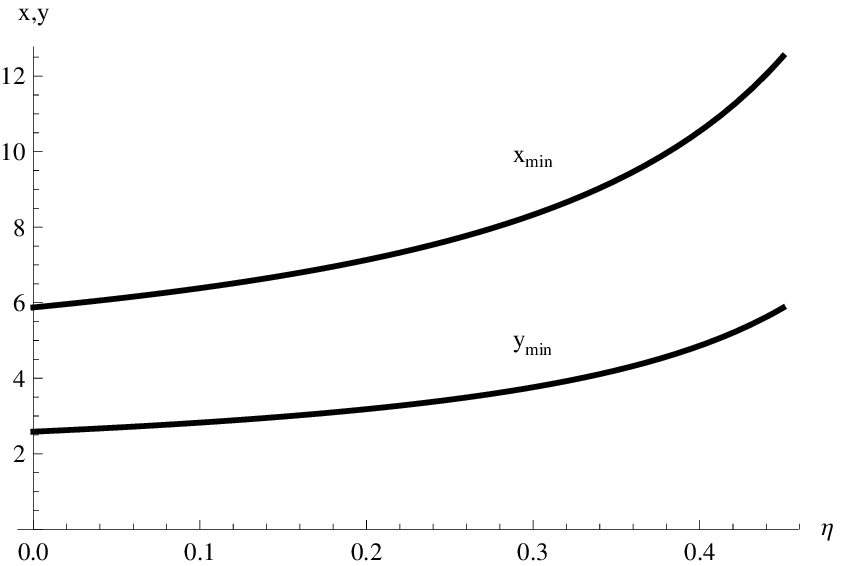}

Fig. 4. Graphs $x_{min}(\eta)$, $y_{min}(\eta)$,.
\end{center}

Fig.3 suggests that $E_{min}(\eta=0)=-0.440636\alpha^2$
yields the lowest estimation of the
bipolaron ground state energy as compared to all those
obtained earlier by variational method. Horizontal lines
in Fig.3 correspond to the energies: $E_1=-0.217\alpha^2$
and $E_2=-0.2515\alpha^2$, where $E_1=2E_{p_1}$, $E_{p_1}$ is
the Pekar polaron ground state energy \eqref{1.3};
$E_2=2E_{p_2}$ where $E_{p_2}$
is the ground
state energy of a translation-invariant polaron \eqref{5.14}.
Intersection of these lines with the curve $E_{min}(\eta)$ yields the
critical values of the parameters $\eta=\eta_{c_1}=0.3325$
and $\eta=\eta_{c_2}=0.289$. For $\eta>\eta_{c_2}$,
the bipolaron decays into two translation-invariant polarons,
for $\eta>\eta_{c_1}$, it breaks down into Pekar polarons.
The values of minimizing parameters $x_{min}$ and $y_{min}$ for these
values of $\eta$  are $x_{min}(0)=5.87561$, $y_{min}(0)=2.58537$,
$x_{min}(0.289)=8.16266$, $y_{min}(0.289)=3.68098$,
$x_{min}(0.3325)=8.88739$, $y_{min}(0.3325)=4.03682$.

The critical value of the electron-phonon coupling constant $\alpha$, determined from comparison of the energy expressions in the weak coupling limit (doubled energy of a weak coupling polaron: $E=-2\alpha\hbar\omega_0$) and in the strong coupling limit ($E=-0.440636\alpha^2\hbar\omega_0$),
at which the translation-invariant bipolaron is formed is
equal to $\alpha_c\approx4.54$, being the lowest estimate obtained by
variational method.
It should be emphasized that this value is conventional. Hamiltonian \eqref{6.4} coincides in structure with one-electron Hamiltonian \eqref{2.3}, therefore, as in the case of a polaron, the bipolaron energy, by \cite{18}, is an analytical function of $\alpha$. For this reason at the point $\alpha=\alpha_c$ the bipolaron energy does not have any peculiarities and the bipolaron state exists over the whole range of $\alpha$ and $\eta$ variation: $0<\alpha<\infty$; $0<\eta<1-1/2\sqrt{2}$, for which $E<0$. To solve the problem of the existence of $\alpha_c$ value at which the bipolaron state can decay into individual polarons, one should perform calculations for the case of intermediate coupling. In particular, a scenario is possible when the bipolaron energy for some values of $\eta$ will be lower than the energy of two individual polarons for all values of $\alpha$, i.e. the bipolaron state exists always.
Notice, that for the derived state of
the translation-invariant bipolaron, the virial theorem
holds to a high precision.

The problems of arising high-temperature superconductivity (HTSC)
and explaning this phenomenon by formation of bipolaron states
were dealt with in a number of papers and reviews \cite{7},
\cite{8}, \cite{55}, \cite{56}.
In these works the existence of HTSC is explained by Bose condensation
of a bipolaron gas. The temperature of Bose condensation
$T_0=3.31\hbar^2n_0^{2/3}/k_B m_{bp}$, which
is believed to be equal to the critical temperature of a superconducting
transition  $T_c$ for $m_{bp}\approx10m$,
depending on the bipolarons concentration $n_0$,
varies in a wide range from $T_0=3K$ at $n=10^{18}cm^{-3}$
to $T_0\approx300K$ at  $n\approx10^{21}cm^{-3}$.
In the latter case the
bipolarons concentration is so high that for a bipolaron gas
as well as for Cooper pairs, the compound character of a bipolaron
when it ceases to behave like an individual particle should show up.
In the case of still higher concentrations a bipolaron should decay
into two polarons. According to \eqref{6.6} the characteristic size of
a bipolaron state is equal to $\ell$ and in dimension units is written as:
$\ell_{corr}=\hbar^2\tilde{\varepsilon}x(\eta)/me^2$.
Here $\ell_{corr}$ has the meaning of a correlation length. The dependence
$x(\eta)$ is given by Fig.4. Fig.4 suggests that over the whole range of $\eta$
variation where the bipolaron state is stable the value of $x$
changes only slightly: from $x(\eta=0)\approx6$  to $x(\eta=0.289)\approx8$.
Hence, even for $\eta=\eta_{c}$,
the critical value of the concentration at which the bipolaron’s
multipiece character is noticeable is of the order of $n_c\cong10^{21}cm^{-3}$.
This result testifies that a bipolaron mechanism of HTSC
can occur in copper oxides.

\section{Ground state functional. Tulub ansatz.}

To diagonalize quadratic form \eqref{3.4} we can use Bogolyubov-Tyablikov
transformation \cite{57}. Let us write $\alpha_k$ for the operators of physical
particles in which $\mathcal{H}_0$ is a diagonal operator.

Let us diagonalize the quadratic form with the use of the transformation:
\begin{equation}\label{7.1}
    \begin{split}
    a_k&=\sum_{k'}M_{1kk'}\alpha_{k'}+\sum_{k'}M_{2kk'}^{*}\alpha_{k'}^{+}\,,\\
    a_k^{+}&=\sum_{k'}M_{1kk'}^{*}\alpha_{k'}^{+}+\sum_{k'}M_{2kk'}\alpha_{k'}\,,
\end{split}
\end{equation}
so that the equalities:
\begin{equation}\label{7.2}
    [a_k,a_{k'}^{+}]=[\alpha_k,\alpha_{k'}^{+}]=\delta_{kk'}\,,\ \
    \ [H_0,\alpha_k^{+}]=\omega_k\alpha_k^{+}\,.
\end{equation}
be fulfilled.

Relying on the properties of unitary transformation \eqref{7.1} we have:
\begin{equation}\label{7.3}
    \begin{split}
    M_2M_1^{+}&=M_1^{*}M_2^{T}\,,\\
    (M_1^{+})^{-1}&=M_1-M_2^{*}(M_1^{*})^{-1}M_2\,.
\end{split}
\end{equation}
With the use of \eqref{7.3} transformation of the operators reciprocal to \eqref{7.1} takes on the form:
\begin{equation}\label{7.4}
    \begin{split}
    \alpha_k&=\sum_{k'}M_{1kk'}^{*}a_{k'}-\sum_{k'}M_{2kk'}^{*}a_{k'}^{+}\,,\\
    \alpha_k^{+}&=\sum_{k'}M_{1kk'}a_{k'}^{+}-\sum_{k'}M_{2kk'}a_{k'}\,.
\end{split}
\end{equation}
According to \cite{27}, \cite{28} matrices $M_1$ and $M_2$ are:
\begin{equation}\label{7.5}
    (M_{1,2})_{kk'}=\frac{1}{2}(\omega_k\omega_{k'})^{-1/2}(\omega_k\pm\omega_{k'})\left[\delta(k-k')+(\vec{k}\vec{k'})f_kf_{k'}
    \frac{2(\omega_k\omega_{k'})^{1/2}}{(\omega_{k'}^2-\omega_k^2\pm
    i\varepsilon)D_{\pm}(\omega_k^2)}\right]\,,
\end{equation}
\begin{equation*}
    D_{\pm}(\omega_p^2)=1+\frac{1}{3\pi^2}\int\limits_{0}^{\infty}\frac{f_k^2k^4\omega_k}{\omega_k^2-\omega_p^2\mp i\varepsilon}dk\,,
\end{equation*}
where the superscript sign in the right-hand side of \eqref{7.5}
 refers to $M_1$ and the subscript sign– to $M_2$.
As a result of diagonalization quadratic form \eqref{3.4} changes to:
\begin{equation}\label{7.6}
    \mathcal{H}_0=\Delta E +\sum_{k}\nu_k\alpha_k^{+}\alpha_k\,.
\end{equation}
Functional of the ground state $\Lambda_0|0\rangle$  is chosen from the condition:
\begin{equation}\label{7.7}
    \alpha_k\Lambda_0|0\rangle=0\,.
\end{equation}
The explicit form of functional $\Lambda_0$ is conveniently derived if we
use Fock representation for operators $a_k$ and $a_k^{+}$
\cite{58}, \cite{59} which
associates operator $a_k^{+}$ with some $c$-number $\bar{a}_k$ and operator $a_k$
with operator $d/d\bar{a}_k$. Then, with the use of \eqref{7.4} condition \eqref{7.7}
takes on the form:
\begin{equation}\label{7.8}
    \left(\sum_{k'}M_{1kk'}^{*}\frac{d}{d\bar{a}_{k'}}-\sum_{k'}M_{2kk'}^{*}\bar{a}_{k'}\right)\Lambda|0\rangle=0\,.
\end{equation}
It is easy to verify by direct substitution in   \eqref{7.8}
that solution of equation \eqref{7.8} is written as:
\begin{equation}\label{7.9}
    \Lambda_0=C\exp\left\{\frac{1}{2}\sum_{k,k'}a_k^{+}A_{kk'}a_{k'}^{+}\right\}\,,
\end{equation}
were $C$  is a constant. To this end it is sufficient to return in \eqref{7.9}
to quantities $\bar{a}_k$ instead of $a_k^{+}$. Matrix $A$ satisfies the conditions:
\begin{equation}\label{7.10}
    A=M_2^{*}(M_1^{*})^{-1}\,,\ \ A=A^{T}\,.
\end{equation}
Hence, the ground state energy corresponding to functional $\Lambda_0$ is equal to:
\begin{equation}\label{7.11}
    \langle0|\Lambda_0^{+}\mathcal{H}_0\Lambda_0|0\rangle=\Delta E\,.
\end{equation}
In Appendix 1 we show that
$\langle0|\Lambda_0^{+}H_1\Lambda_0|0\rangle\equiv0$.

From \eqref{7.9}, \eqref{2.2}, \eqref{2.5} it follows that the wave
function of the polaron ground state $|\psi\rangle_{p}$ has the form:
\begin{equation}\label{7.12}
    |\psi\rangle_{p}=Ce^{-\frac{i}{\hbar}\sum\limits_{k}\hbar\vec{k}a_k^{+}a_k\vec{r}}e^{\sum\limits_{k}f_k(a_k^{+}-a_k)}\Lambda_0|0\rangle\,.
\end{equation}
Accordingly the bipolaron wave function $|\psi\rangle_{bp}$, with regard to \eqref{5.3}, \eqref{5.4} is:
\begin{equation}\label{7.13}
    |\psi\rangle_{bp}=C\psi(r)e^{-\frac{i}{\hbar}\sum\limits_{k}\hbar\vec{k}a_k^{+}a_k\vec{R}}e^{\sum\limits_{k}f_k(a_k^{+}-a_k)}\Lambda_0|0\rangle\,.
\end{equation}
Formula \eqref{7.12}, \eqref{7.13} imply that the wave functions of a polaron
and bipolaron are delocalized over the whole space and cannot
be presented as an ansatz \eqref{1.2}.

From formulae \eqref{7.12}, \eqref{7.13} it follows that the reason why the attempt
of Lee, Low and Pines \cite{31} to investigate the polaron ground state energy
over the whole range of $\alpha$ variation failed was an improper choice of probe
function \eqref{2.4} which lacks the multiplier corresponding to the functional $\Lambda_0$.

However, it should be stressed that notwithstanding a radical improvement
of the wave function achieved by introducing the multiplier $\Lambda_0$
in Lee,  Low, Pines function
enables one to take account of both weak and strong coupling, the results
obtained by its application are not exact. The fact that Tulub function
is an ansatz follows from its properties:
\begin{equation}\label{7.14}
    \langle0|\Lambda_0^{+}H_1\Lambda_0|0\rangle=0\,,\ \ E=\langle0|\Lambda_0^{+}H_0\Lambda_0|0\rangle\,,\ \
    H_0\Lambda_0|0\rangle=E\Lambda_0|0\rangle\,.
\end{equation}
Being an ansatz, Tulub’s solution presents a solution of the polaron problem
in a specific class of functions having the structure of $\Lambda_0|0\rangle$.
That Tulub ansatz
is not an exact solution of the problem follows at least from the fact that
the use of expression \eqref{3.20} alone for calculation of the energy,
for example, in the case of weak coupling yields for $E$:
${E=-\alpha-\frac{1}{6}\left(\frac{1}{2}-\frac{4}{3\pi}\right)\alpha^2}$ \cite{27}.
To get an
exact coefficient at $\alpha^2$ in the expansion of the energy in powers of $\alpha$ \eqref{1.1}
we should take into account the contribution of Hamiltonian $H_1$, as the
perturbation theory suggests \cite{34}.

The fact that wave functions \eqref{7.12},
\eqref{7.13} are delocalized has a lot of
important consequences which will be discussed in Section 9.

\section{Discussion of the completeness of Tulub theory.}

In \cite{60}, \cite{61} a question was raised as to whether Tulub theory
\cite{27}, \cite{28} is complete. Arguments of \cite{60}, \cite{61} are based on
the work by Porsch and R\"{o}seler \cite{33} which reproduces the
results of Tulub theory. However, in the last section of
their paper Porsch and R\"{o}seler investigate what will happen
if the infinite integration limit in Tulub theory changes
for a finite limit and then passes on to the infinite one.
Surprisingly, it was found that in this case in parallel
with cutting of integration to phonon wave vectors in the
functional of the polaron total energy one should augment
the latter with the addition $\delta E^{PR}$ which input will not disappear
if the upper limit tends to infinity \cite{33},
\cite{61}.
Relying on this result the authors of \cite{60},
\cite{61} concluded
that Tulub did not take this addition into account and
therefore his theory is incomplete.

To resolve this paradox let us consider the function $\Delta(s)$ determined
by formula \eqref{3.14} (accordingly, \eqref{3.19} in continuous case).
As formulae \eqref{3.14}, \eqref{3.19} imply, zeros in this function contribute
into "polaron recoil"\ energy $\Delta E$ given by \eqref{3.16} and, according
to \eqref{3.15} are found from the solution of the equation:
\begin{equation}\label{8.1}
    1=\frac{2}{3}\sum_{k}\frac{k^2f_k^2\omega_k}{s-\omega_k^2}\,.
\end{equation}
If the cutoff is absent in the sum on the right-hand side of
equation \eqref{8.1}, then the solution of this equation yields
a spectrum of $s$ values determined by frequencies $\nu_{k_i}$ lying
between neighboring values of $\omega_{k_i}$ and $\omega_{k_{i+1}}$
for all the wave vectors $k_i$.
These frequencies determine the value of the polaron recoil energy:
\begin{equation}\label{8.2}
    \Delta E=\frac{1}{2}\sum_{k_i}(\nu_{k_i}-\omega_{k_i})\,.
\end{equation}
Let us see what happens with the contribution of frequencies $\nu_{k_i}$
into $\Delta E$ in the region of the wave vectors $k$ where $f_k$ vanishes
but nowhere becomes exactly zero. From \eqref{8.1} it follows that
as $f_k\to0$, solutions of equation \eqref{8.1} will tend to $\omega_{k_i}$: $\nu_{k_i}\to\omega_{k_i}$.
Accordingly, the contribution of the wave vectors region into $\Delta E$,
where $f_k\to0$, will also tend to zero.

In particular, if we introduce a certain $k^0$ such that in the
region $k>k^0$ the values of $f_k$ are small, we will express
$\Delta E$ in the form:

\begin{equation}\label{8.3}
    \Delta
    E=\frac{1}{2}\sum_{k_i\leq k^0}(\nu_{k_i}-\omega_{k_i})\,,
\end{equation}
which does not contain any additional terms. To draw a parallel
with Tulub approach, there we could put the upper limit $k^0$ ,
but no additional terms would appear.

For example, if in an attempt to investigate the minimum
of Tulub functional \eqref{3.20}, \eqref{5.7} we choose the probe function
$f_k$ not containing a cutoff in the form \cite{45}:
\begin{equation}\label{8.4}
    \begin{split}
    f_k&=-V_k\exp(-k^2/2a^2(k))\,,\\
    a(k)&=\frac{a}{2}\left[1+\tah\left(\frac{k_b-k}{a}\right)\right]\,,
\end{split}
\end{equation}
where $a$ is a parameter of Tulub probe function \eqref{5.9},
$k_b$ satisfies the condition $a\ll k_b \ll k_{oc}$,
$k_{oc}=a\sqrt[4]{3\lambda/4}$ is the value of the wave
vector for which Tulub integral \eqref{5.12} has a maximum
\cite{28}, \cite{62}, then with the use of \eqref{8.4} in the limit $a\to\infty$ ,
Tulub integral $q(1/\lambda)$ will be written as:
\begin{equation}\label{8.5}
    q\left(\frac{1}{\lambda}\right)\approx5.75+6\left(\frac{a}{k_b}\right)^3\exp\left(-\frac{k_b^2}{a^2}\right)\,.
\end{equation}
The second term in the right-hand side of \eqref{8.5} vanishes as $k_b/a\to\infty$
and we get, as we might expect, Tulub’s result: $q(1/\lambda)\approx5.75$.

Equation \eqref{8.1}, however, has a peculiarity. Even in the case of
a continuous spectrum, for $f_k=0$, if $k>k^0$ it has an isolated solution
$\nu_{k^0}$ which differs from the maximum frequency $\omega_{k^0}$ by a finite value.
This isolated solution leads to an additional contribution into $\Delta E$:
\begin{equation}\label{8.6}
    \begin{split}
    \Delta E&=\frac{1}{2}\sum_{k_i< k^0}(\nu_{k_i}-\omega_{k_i})+\delta E^{PR}\,,\\
    \delta E^{PR}&=\frac{3}{2}(\nu_{k^0}-\omega_{k^0})\,,
\end{split}
\end{equation}
where $\nu_{k^0}$ has the meaning of "plasma frequency"\ \cite{33}.
Hence, here a continuous transition from the case of $f_k\to0$
for $k>k^0$ to the case of $f_k=0$ for $k>k^0$ is absent.
As is shown by direct calculation \cite{63}, of the contribution
of the term with "plasma frequency"\ $\delta E^{PR}$ into \eqref{8.6}, even for $k^0\to\infty$,
Porsch and R\"{o}seler theory does not transform itself
into Tulub theory.

In Tulub theory we choose such $f_k$ which lead to the minimum
of the functional of the polaron total energy. In particular,
the choice of the probe function in the form \eqref{8.4} provides
the absence of a contribution from "plasma frequency"\ into
the total energy and in actual calculations one can choose
a cutoff $f_k$ without introducing any additional terms
in Tulub functional \cite{62}, \cite{63}.

To sum up, critical remarks in \cite{60}, \cite{61} are inadequate.
Their inadequacy was demonstrated in papers \cite{62}, \cite{63} and
in work \cite{45} reproduced here. At the present time Tulub
theory and the results obtained on its basis \cite{29}, \cite{43}-\cite{45}
give no rise to doubt.

\section{Consequences of the existence of translation-invariant polarons and bipolarons.}

According to the results obtained, the ground state of a
TI-polaron is a delocalized state of the electron-phonon
system: the probabilities of electron’s occurrence at
any point of the space are similar. The explicit form
of the wave function of the ground state is presented in \S7.
Both the electron density and the amplitudes of phonon modes
(corresponding to renormalized by interaction frequencies $\nu_{q_i}$)
are delocalized.

It should be noted that according to \eqref{3.15} renormalized phonon
frequencies $\nu_{q_i}$ in the case of a TI-polaron have higher energies
than non-renormalized frequencies
of optical phonons and, therefore, higher energies than
non-renormalized frequencies of a polaron with spontaneously
broken symmetry \cite{64}.
This holds out a hope to find such phonon modes in experiments on light combination scattering and optical absorption. According to \cite{64}, if a polaron (bipolaron) is bound on the Coulomb center, i.e. it forms an $F$-center ($F'$-center), then all renormalized local phonon frequencies $\omega_n$ have lower energies than the frequency of optical phonon $\omega_0$ does. This fact also makes easier experimental validation of the occurrence of delocalized TI-phonon modes with $\nu_{q_i} > \omega_0$.

The concept of a polaron potential well (formed by local phonons
\cite{64}) in which an electron is localized, i.e. the self-trapped
state is lacking in the translation-invariant theory.
Accordingly, the induced polarization charge of the TI-polaron
is equal to zero. Polaron’s lacking a localized "phonon environment"\
suggests that its effective mass is not very much different
from that of an electron. The ground state energy of a TI-polaron
is lower than that of Pekar polaron and is given by formula \eqref{5.14}
(for Pekar polaron the energy is determined by \eqref{1.3}).

Hence, for zero total momentum of a polaron, there is an energy
gap between the TI-polaron state and the Pekar one (i.e. the state
with broken translation invariance). The TI-polaron is a
structureless particle (the results of investigations of the Pekar
polaron structure are summed up in \cite{64}).

According to the translation-invariant polaron theory, the terms
"large-radius polaron"\ (LRP) and "small-radius polaron"\ (SRP)
are relative, since in both cases the electron state is delocalized
over the crystal. The difference between the LRP and SRP in the
translation-invariant theory lies in the fact that for the LRP
the inequality $k_{char}a<\pi$  is fulfilled, while for the SRP $k_{char}a>\pi$
holds, where $a$ is the lattice constant and $k_{char}$ is a characteristic
value of the phonon wave vectors making the main contribution into
the polaron energy. This statement is valid not only for
Pekar-Froehlich polaron, but for the whole class of polarons
whose coupling constant is independent of the electron wave vector,
such as Holstein polaron, for example. For polarons whose coupling
constant depends on the electron wave vector, these criteria may
not hold (as is the case with Su-Schreiffer-Heeger model,
for example \cite{65}).

These properties of TI-polarons determine their physical
characteristics which are qualitatively different from those
of Pekar polarons. When a crystal has minor local disruptions,
the TI-polaron remains delocalized. For example, in an
ionic crystal containing vacancies, delocalized polaron
states will form $F$-centers only at a certain critical value
of the static dielectric constant $\varepsilon_{0c}$.
For $\varepsilon_0>\varepsilon_{0c}$ , a crystal will
have delocalized TI-polarons and free vacancies. For $\varepsilon_0=\varepsilon_{0c}$,
a transition from the delocalized state to that localized
on vacancies (collapse of the wave function) will take place.
Such behavior of translation-invariant polarons is qualitatively
different from that of Pekar polarons which are localized
on the vacancies at any value of $\varepsilon_0$. This fact accounts for,
in particular, why free Pekar polaron does not demonstrate
absorption (i.e. structure), since in this case the
translation-invariant polaron is realized.
Absorption is observed only when a bound Pekar polaron,
i.e. $F$-center is formed. These statements are also supported
by a set of recent papers where Holstein polaron
is considered \cite{66}-\cite{68}. The approach presented is generalized by the author to the case of Holstein polaron in \cite{69n-Lakhno}.

Notice that the physics of only free strong-coupling polarons
needs to be changed. The overwhelming majority of results on
the physics of strong-coupling polarons has been obtained for
bound (on vacancies or lattice disruptions) polaron states of
Pekar type and do not require any revision.

Taking account of translation invariance in the case of a polaron
leads to a minor change in the assessment of the ground state,
however leads to qualitatively different visions
of the properties of this state. In paper \cite{28}, in the section
devoted to scattering of a TI-polaron, Tulub shows that as
the constant of electron-phonon coupling increases up to
a certain critical value, scattering of an electron on optical
phonons turns to zero. Hence, when the coupling constants exceed
a critical value a polaron becomes superconducting.
Though in ionic crystals the main mechanism of electron scattering
is scattering on optical phonons \cite{69}, it might appear that
the contribution of acoustic phonons should also be taken
into account in this case. However, as it follows from the
law of conservation of energy and momentum, a TI-polaron
will scatter on acoustic phonons only if its velocity
exceeds that of sound \cite{70}.

As distinct from polarons, TI-bipolarons have much greater
binding energy. This leads to some important physical consequences.
In particular, when a crystal has minor local disruptions,
a TI-bipolaron will be delocalized. Thus, in an ionic crystal
with lattice vacancies, formation of $F'$-centers by delocalized
bipolarons will take place only at a certain critical value
of the static dielectric constant $\varepsilon_{0c_1}$.
For $\varepsilon_0>\varepsilon_{0c_1}$, the crystal
will contain delocalized TI-bipolarons and free vacancies.
In the case of $\varepsilon_0=\varepsilon_{0c_1}$
TI-bipolarons will pass on from the
delocalized state to that localized on vacancies, i.e.
to $F'$-center. Such behavior of TI-bipolarons is qualitatively
different from the behavior of polarons with spontaneously
broken symmetry of Pekar type \cite{8}, which are localized
on the vacancies at any value of $\varepsilon_0$.

The fundamental difference between TI-bipolarons and bipolarons with spontaneously broken symmetry is that the former are not separable while the latter are separable. This is due to the fact that in the case of bipolarons with spontaneously broken symmetry interaction between electrons and polarization has the form: $\Phi\left(\vec{r}_1,\vec{r}_2\right) = F(\vec{r}_1) + F(\vec{r}_2)$. For $|\vec{r}_1-\vec{r}_2| \gg R$, where $R$ is the bipolaron radius, bipolaron equations separate into two independent polaron equations. This fact enables us to treat a bipolaron state with spontaneously broken symmetry as a bound state of two polarons \cite{8}. In the case of TI-bipolarons: $\Phi(\vec{r}_1,\vec{r}_2) = \Phi(\vec{r}_1-\vec{r}_2)$. In this case, splitting of the bipolaron interaction functional into the functionals of interaction of individual polarons is impossible for any $|\vec{r}_1-\vec{r}_2|$ and treatment of TI-bipolarons as composite states becomes invalid. This conclusion corresponds to modern ideas that a quantum-mechanical system cannot be separated into independent subsystems \cite{71n-Grib}.

For zero total momentum of a bipolaron, TI-bipolarons,
being delocalized, will be separated from those with broken
translation invariance by an energy gap. As with TI-polarons,
in the case when the coupling constant exceeds a certain
critical value, TI-bipolarons become superconducting.
As is known, interpretation of the high-temperature superconductivity
relying on the bipolaron mechanism of Bose-condensation runs into
a problem associated with a great mass of bipolarons and,
consequently low temperature of Bose-condensation.
The possibility of smallness of TI-bipolarons’ mass resolves
this problem. It should be stressed that the above-mentioned
properties of translation-invariant bipolarons impart them
superconducting properties even in the absence of Bose-condensation,
while the great binding energy of bipolarons substantiates
the superconductivity scenario even in badly defect crystals.

\section{Conclusive remarks.}

At the present time Tulub theory and quantitative results obtained
on its basis give no rise to doubt. The quantum field theory
under consideration is nonperturbative and can reproduce not
only the limits of weak and strong coupling but also the
case of intermediate coupling.

One of the most effective methods for calculating polarons and
bipolarons in the case of intermediate coupling is integration
over trajectories \cite{71}. Unless properly modified, this approach
is not translation-invariant since in this method the main
contribution into the energy levels is given by classical
solutions (i.e. extrema points of the exponent of classical action,
involved in the path integral). However, such solutions,
in view of translation invariance, are not isolated stationary
points, but belong to a continuous family of classical
solutions obtained as a result of action of the translation
operator on the initial classical solution. Accordingly,
the stationary phase approximation is inapplicable in the
translation-invariant system. In the quantum field theory
some approaches are developed for restoring translation invariance.
They are based on introducing collective coordinates into
the functional integral \cite{72}. However, they have not been
used in the polaron theory as yet. Therefore it is not surprising
that the method of integrals over trajectories employed
in the plaron theory yields a result coinciding with
the semiclassical theory of the strong coupling polaron \cite{73}.

Recently in the polaron theory a powerful computational method,
namely Monte-Carlo technique has been developed \cite{74}, \cite{75}.
This procedure, being only a calculation tool, cannot reproduce
the results of Tulub ansatz unless properly modified.
As for Monte-Carlo diagram technique, the obstacle
to checking Tulub ansatz in the strong coupling limit is presented
by the necessity to calculate diagrams of very high order.

To sum up, Pekar ansatz \eqref{1.2} provides an original assumption
of the form of the solution which was confirmed in the course
of numerous examinations. For nearly eighty-year history
of the polaron theory development (if dating from Landau short paper \cite{76})
ansatz \eqref{1.2} has been recognized to be an asymptotically
exact solution of the polaron problem in the strong coupling limit.

Tulub ansatz (\S7) provides another assumption of the form of
the solution whose structure is determined by the form of
the function $\Lambda_0|0\rangle$. In terms of this assumption Tulub solution
is also asymptotically exact. Since Tulub solution yields
a lower energy value for a polaron, from the variational standpoint,
preference should be given to Tulub ansatz.

Hence, polaron theory in no way can be considered to be
complete. In the framework of Tulub ansatz great work
is to be done to revise many concepts (such as superconductivity)
in condensed matter physics. Extension of the application area
of Tulub ansatz to other divisions of the quantum field theory
can lead to radical revision of many results which nowadays seem doubtless and vice-versa.
Thus, for example, non-separability of a bipolaron state in the polaron model of quarks \cite{81n-Iwao} (the role of phonons in \cite{81n-Iwao} is played by a gluon field) provides a natural explanation of their confinement. In paper \cite{69n-Lakhno} it is noted, that in TI-theory there is no need to use Higgs mechanism of spontaneous symmetry breaking to get the elementary particles masses.

The author expresses sincere gratitude to Prof. A.V. Tulub for numerous
discussions and valuable advices. The author is also grateful
to N. I. Kashirina for valuable discussions.

Various aspects of the problems considered here have also
been talked over with V.A. Osipov, E.A. Kochetov, S.I. Vinitsky
to whom the author also renders his thanks.

The work was done with the support from the RFBR, Project N 13-07-00256.

\section*{Appendix 1.}

Hamiltonian $H_1$ involved in \eqref{3.2} has the form:
\begin{equation*}
    H_1=\sum_{k}(V_k+f_k\hbar\omega_k)(a_k+a_k^{+})+\sum_{k,k'}\frac{\vec{k}\vec{k'}}{m}f_{k'}(a_k^{+}a_ka_{k'}+a_k^{+}a_{k'}^{+}a_k)+
    \frac{1}{2m}\sum_{k,k'}\vec{k}\vec{k'}a_k^{+}a_{k'}^{+}a_ka_{k'}\,,\eqno{(A1.1)}
\end{equation*}
where $\hbar\omega_k$  is given by expression \eqref{3.5}.
Let us apply the operator
$H_1$ to functional $\Lambda_0$ \eqref{7.9}. We will show that $\langle0|\Lambda_0^{+}H_1\Lambda_0|0\rangle=0$.
Indeed, the action of $\Lambda_0$ on  $H_1$  terms containing
an odd number of operators in $H_1$ (i.e. the first
and second terms in $H_1$) will always contain an odd
number of terms and mathematical expectation for these
terms will tend to zero.

Let us consider mathematical expectation for the last term in $H_1$:
\begin{equation*}
    \langle0|\Lambda_0^{+}\sum_{k,k'}\vec{k}\vec{k'}a_k^{+}a_{k'}^{+}a_ka_{k'}\Lambda_0|0\rangle\,.\eqno{(A1.2)}
\end{equation*}
The function
$\langle0|\Lambda_0^{+}a_k^{+}a_{k'}^{+}a_ka_{k'}\Lambda_0|0\rangle$
represents the norm of vector $a_ka_{k'}\Lambda_0|0\rangle$
and will be positively
defined for all  $k$  and $k'$. If we replace $\vec{k}\rightarrow-\vec{k}$ in (A1.2)
than the whole expression will change the sign and,
therefore, (A1.2) is also equal to zero. Hence
$\langle0|\Lambda_0^{+}H_1\Lambda_0|0\rangle=0$.

\section*{Appendix 2.}

Let us show that \eqref{5.4}, \eqref{5.5} follow from \eqref{5.1},
\eqref{5.2}.
To this end notice that analytical properties of $D(s)$
pointed out in \cite{28} straightforwardly follow from \eqref{3.19}.
Indeed, the pole of
$D(s)$ can lie only on the real axis since in view of
positive definedness of $\omega_kk^4f_k^2$ in \eqref{3.19} equation:
\begin{equation*}
    1+\frac{1}{3\pi^2}\int\limits_{0}^{\infty}\frac{\omega_kk^4f_k^2(\omega_k^2-s_0+i\varepsilon)}{(\omega_k^2-s_0)^2+\varepsilon^2}dk=0\,,\eqno{(A2.1)}
\end{equation*}
can be fulfilled only for $\varepsilon=0$. Besides, $D(s)$ is a monotonously
increasing function $s$  since for $s<1$,  $D'(s)>0$, and as $s_0\to\infty$,
$D(s)$ turns to unit. Therefore $D(s)$ cannot have zeros for
$-\infty<s_0<1$. Hence function $(s-1)D(s)$ can be presented in the form:
\begin{equation*}
    \frac{1}{(s-1)D(s)}=\frac{1}{2\pi
    i}\oint\limits_{C+\rho}\frac{ds'}{(s'-s)(s'-1)D(s')}\,,\eqno{(A2.2)}
\end{equation*}
where the contour of integration in Caushy integral $(A2.2)$
is shown in Fig.2. The integrand function in $(A2.2)$ has a
pole for $s'=1$ and a section from  $s'=1$ to $s'\to\infty$.
Performing integration in $(A2.2)$ with respect to the upper
and bottom sides of the crosscut we will get integral
equation \eqref{5.5}.

\section*{Appendix 3.}

Let us perform a detailed calculation of the quantity
$\Delta E$ \eqref{5.7} with the use of probe function \eqref{5.9}.

To this end, to calculate the real part of $D(\omega_p^2)$
involved in \eqref{5.7} we use Sokhotsky’s formula:
\begin{equation*}
    \frac{1}{\omega_k^2-\omega_p^2-i\varepsilon}=\mathcal{P}\frac{1}{\omega_k^2-\omega_p^2}+i\pi\delta(\omega_k^2-\omega_p^2)\,,
\end{equation*}
\begin{equation*}
    \Rer
    D(\omega_p^2)=1+\frac{1}{3\pi^2}\int\limits_{0}^{\infty}f_k^2k^4\mathcal{P}\frac{\omega_k}{\omega_k^2-\omega_p^2}dk\,.
\end{equation*}
It is convenient to present $\Rer D$ in the form:
\begin{equation*}
    \Rer D=1+I_1+I_2\,,
\end{equation*}
\begin{equation*}
    I_1=\frac{1}{3\pi^2}\int\limits_{0}^{\infty}f_k^2k^4\frac{dk}{\omega_k+\omega_p}\,,\
    \
    I_2=\mathcal{P}\frac{\omega_p}{3\pi^2}\int\limits_{0}^{\infty}\frac{f_k^2k^4dk}{(\omega_k-\omega_p)(\omega_k+\omega_p)}\,.
\end{equation*}
Substituting $f_k$ in the form of \eqref{5.9} into these expressions we present $I_1$ as:
\begin{equation*}
    I_1=\frac{8\alpha}{3\sqrt{2}\pi}\int\limits_{0}^{\infty}e^{-k^2/a^2}dk-\frac{8\alpha(p^2+4)}{3\sqrt{2}\pi}\int\limits_{0}^{\infty}
    \frac{e^{-k^2/a^2}}{k^2+p^2+4}dk\,.
\end{equation*}
Assuming $k/a=\tilde{k}$ in the strong coupling limit ($a\to\infty$) we write for $I_1$:
\begin{equation*}
    I_1=\frac{8\alpha a}{3\sqrt{2}\pi}\left[\frac{\sqrt{\pi}}{2}-\frac{\pi}{2}\tilde{p}e^{\tilde{p}^2}
    \left(1-\frac{2}{\sqrt{\pi}}\int\limits_{0}^{\tilde{p}}e^{-t^2}dt\right)\right]\,.
\end{equation*}
Accordingly, $I_2$  takes on the form:
\begin{equation*}
    I_2=\mathcal{P}\frac{4\alpha\omega_p}{3\pi\sqrt{2}}\int\limits_{0}^{\infty}\frac{e^{-k^2/a^2}k^2dk}{(\omega_k-\omega_p)(\omega_k+\omega_p)\,.}
\end{equation*}
This integral can be expressed as:
\begin{equation*}
    I_2=I_{20}+I_{21}\,,
\end{equation*}
where:
\begin{equation*}
    I_{20}=\frac{16\alpha\omega_p}{3\pi\sqrt{2}}\left(1-\frac{\omega_p-1}{p^2+2}\right)\int\limits_{0}^{\infty}\frac{e^{-k^2/a^2}}{k^2+p^2+4}dk\,,
\end{equation*}
\begin{equation*}
    I_{21}=\frac{16\alpha\omega_p(\omega_p-1)}{3\pi\sqrt{2}(p^2+2)}\mathcal{P}\int\limits_{0}^{\infty}\frac{e^{-k^2/a^2}}{k^2-p^2}dk\,.
\end{equation*}
The integrals involved in $I_{20}$ and $I_{21}$  will be:
\begin{equation*}
    \int\limits_{0}^{\infty}\frac{e^{-k^2/a^2}}{k^2+p^2+4}dk=\frac{1}{a}\left[1-\frac{2}{\sqrt{\pi}}\int\limits_{0}^{\tilde{p}}e^{-t^2}dt\right]
    \frac{\pi}{2}\frac{e^{\tilde{p}^2}}{\tilde{p}}\,,
\end{equation*}
\begin{equation*}
    \mathcal{P}\int\limits_{0}^{\infty}\frac{e^{-k^2/a^2}}{k^2-p^2}dk=-\frac{\sqrt{\pi}}{a}\frac{e^{-\tilde{p}^2}}{\tilde{p}}
    \int\limits_{0}^{\tilde{p}}e^{t^2}dt\,.
\end{equation*}
As a result,  $I_2$ has the form:
\begin{equation*}
    I_2=\frac{2}{3}\frac{\alpha
    a\tilde{p}}{\sqrt{2}}e^{\tilde{p}^2}\left[1-\frac{2}{\sqrt{\pi}}\int\limits_{0}^{\tilde{p}}e^{-t^2}dt\right]-
    \frac{4\alpha
    a\tilde{p}}{3\sqrt{2\pi}}e^{-\tilde{p}^2}\int\limits_{0}^{\tilde{p}}e^{t^2}dt\,.
\end{equation*}
Finally,  $\Rer D$ will be written as:
\begin{equation*}
    \Rer D=1+\frac{4\alpha a}{3\sqrt{2\pi}}\left(1
    -\tilde{p}e^{\tilde{p}^2}\int\limits_{\tilde{p}}^{\infty}e^{-t^2}dt-\tilde{p}e^{-\tilde{p}^2}\int\limits_{0}^{\tilde{p}}e^{t^2}dt\right)\,.
\end{equation*}
This result reproduces the quantity given by formula
\eqref{5.10}. For the imaginary part $\Imi D$ by Sokhotsky’s formula,
we get:
\begin{equation*}
    \Imi
    D=\frac{1}{3\pi}\int\limits_{0}^{\infty}f_k^2k^4\omega_k\delta(\omega_k^2-\omega_p^2)dk=\frac{1}{6\pi}f_p^2p^3\,.
\end{equation*}
As a result, $|D(\omega_k^2)|$ is expressed as:
\begin{equation*}
    |D|^2=(\Rer D)^2+(\Imi D)^2=\frac{2}{9}\alpha^2 a^2\left[e^{-2\tilde{p}^2}\tilde{p}^2+\frac{8}{2\pi}\left(1
    -\tilde{p}\int\limits_{\tilde{p}}^{\infty}e^{-t^2}dt-\tilde{p}e^{-\tilde{p}^2}\int\limits_{0}^{\tilde{p}}e^{t^2}dt\right)^2\right]\,.
\end{equation*}
The first term in formula \eqref{5.7} is easily calculated to be:
\begin{equation*}
    \frac{1}{4\pi^2}\int\limits_{0}^{\infty}\frac{k^4f_k^2}{(1+Q)}dk=\frac{3}{16}a^2\,.
\end{equation*}
In calculating the second term in \eqref{5.7} we will separate out integral $I_p$ in it:
\begin{equation*}
    I_p=\int\limits_{0}^{\infty}e^{-k^2/a^2}\frac{k^2(\omega_k\omega_p+\omega_k(\omega_k+\omega_p)+1)}{(\omega_k+\omega_p)^2}dk\,.
\end{equation*}
As $a\to\infty$, it is equal to:
\begin{equation*}
    I_p=a^3\frac{\sqrt{\pi}}{4}\left(1-\tilde{p}^3e^{\tilde{p}^2}\int\limits_{\tilde{p}}^{\infty}e^{-t^2}dt(2+4\tilde{p}^2)+2\tilde{p}^4\right)=
    \frac{a^3\sqrt{\pi}}{4}(1-\Omega(\tilde{p}))\,,
\end{equation*}
where:
\begin{equation*}
    \Omega(\tilde{p})=2\tilde{p}\left\{(1+2\tilde{p}^2)\tilde{p}e^{\tilde{p}^2}\int\limits_{\tilde{p}}^{\infty}e^{-t^2}dt-\tilde{p}^2\right\}\,,
\end{equation*}
which corresponds to the expression $\Omega(y)$  in \eqref{5.12}
As a result, the second term in formula \eqref{5.7} will be:
\begin{equation*}
    \frac{1}{12\pi^4}\frac{4\pi\alpha}{\sqrt{2}}\int\limits_{0}^{\infty}I_pp^4f_p^2\frac{\omega_p}{(\omega_p^2-1)|D(\omega_p^2)|^2}dp\,.
\end{equation*}
As $a\to\infty$, this expression takes on the form:
\begin{equation*}
    \frac{\alpha^2a^4}{3\pi\sqrt{\pi}}\int\limits_{0}^{\infty}(1-\Omega(\tilde{p}))\frac{e^{-\tilde{p}^2}}{|D(\omega_{\tilde{p}}^2)|^2}d\tilde{p}=\frac{3}{16}a^2q\,,
\end{equation*}
where $q=q(0)$ is given by expression \eqref{5.12}.
Hence, finally for $\Delta E$ \eqref{5.7} we get:
\begin{equation*}
    \Delta E=\frac{3}{16}a^2(1+q)\,,
\end{equation*}
which corresponds to the first term in the right-hand side of \eqref{5.11}.

\end{document}